\documentclass[sigconf]{acmart}
\renewcommand\footnotetextcopyrightpermission[1]{}
\usepackage{enumitem}
\usepackage{multirow}
\usepackage{caption}
\usepackage{babel}
\usepackage{url}
\usepackage{booktabs}
\usepackage{mathtools}
\usepackage{amsmath}

\usepackage{amssymb,amsfonts,amsthm}
\usepackage{graphicx}
\usepackage{textcomp}
\usepackage{stfloats}
\usepackage{xcolor}
\usepackage{xspace,hyperref}

\hypersetup{
    colorlinks=true,
    urlcolor=blue,
    linkcolor=black,
    citecolor=black
}

\usepackage{xspace}
\usepackage{tikz}

\newcommand{\sys}{\textsc{Stamp}\xspace}
\newcommand{\zm}{$\mathsf{ZoneMinder}$\xspace}
\newcommand{\bsection}[1]{\noindent\textbf{#1. }}
\newcommand*\wcircled[1]{\tikz[baseline=(char.base)]{
    \node[shape=circle,draw,fill=white,text=black,inner sep=0.5pt] (char) {\textbf{#1}};}}

\begin{document}

\title{Inferring Sensitive Events in Multi-Mode Cyber-Physical Systems via Scheduling Side Channels}

\author{Yifan Cai}
\authornote{The work was done when Yifan was a PhD student at the University of Pennsylvania.}
\affiliation{
    \institution{University of British Columbia}
    \city{Vancouver}
    \state{BC}
    \country{Canada}
}

\author{Linh Thi Xuan Phan}
\affiliation{
    \institution{University of Pennsylvania}
    \city{Philadelphia}
    \state{PA}
    \country{United States}
}

\sloppy

\begin{abstract}
Many cyber-physical systems (CPS) are {\em multi-mode}: they switch between operating modes with different tasksets and timing parameters in response to physical events or user activity. We show that these mode changes create a timing side channel: by altering the interference experienced by co-running tasks, they let an unprivileged co-resident observer infer sensitive events, such as a homeowner's arrival or a vehicle's route, from its own execution timing and, on multicore platforms, its core assignment. Existing schedule-based attacks and defenses miss this channel because they assume single-mode systems or target a specific, known victim task. We present \sys, a lightweight attack that needs no knowledge of task parameters or scheduling policy, and works whether or not the system exposes its modes. On a \zm home-security deployment, \sys identifies the modes with 94.7\% accuracy on unseen hardware, using under 0.5\% of one core. On ROS~2 AMCL with the Oxford RobotCar dataset, it identifies which of seven route segments a vehicle traversed with 86.0\% accuracy on unseen platforms.  
A schedule-randomization defense lowers detection recall to 0.53, at which setting only 53.9\% as many jobs finish on time. At higher noise ($\epsilon = 10$), on-time job completion reduces to 14.4\%, while introducing new leakage in 14.1\% of the tasksets. 
\end{abstract}

\maketitle
\fancyhead{}
\section{Introduction}\label{sec:intro}

Real-time cyber-physical systems (CPS), such as autonomous vehicles and smart-home devices, rely on predictable task execution and scheduling to meet their timing constraints. These systems are often \emph{multi-mode}~\cite{RC04,gifford2022multi,DBLP:conf/rtas/GiffordGPZ24}: they switch between operating modes with different tasksets and timing parameters in response to user commands, environmental changes, or reconfigurations. For instance, an autonomous vehicle may adjust its sensing, perception, and planning workloads as it drives, and a home-security system may start analyzing camera streams when its owner leaves. Because mode changes follow physical events, the schedule itself carries information about the physical world.

We show that this information leaks to co-running tasks through a scheduling side channel: a mode change alters the execution patterns of these tasks, shifting when and on which cores they execute. An unprivileged co-resident task can observe these shifts in its own execution timing and core assignment, and use them to infer sensitive events, such as a homeowner's arrival or a vehicle's route. Such inferences enable privacy and physical-security threats, including cyber-enabled burglary~\cite{DBLP:journals/compsec/Hodges21}.

\begin{table}[t]
\centering
\small
\begin{tabular}{llll}
\hline
\textbf{Work}                              & \textbf{Immediate Target}                                                    & \textbf{Mode}     & \textbf{\begin{tabular}[c]{@{}l@{}}Scheduling\\ policy\end{tabular}}       \\ \hline
ScheduLeak~\cite{chen2019novel}            & \begin{tabular}[c]{@{}l@{}}Execution slot\\ of a victim task\end{tabular} & Single            & \begin{tabular}[c]{@{}l@{}}Known\\ fixed-priority\end{tabular}             \\ \hline
DyPS~\cite{chen2020scheduler}              & \begin{tabular}[c]{@{}l@{}}Execution slot\\ of a victim task\end{tabular} & Single            & Known EDF                                                                  \\ \hline
ETPT~\cite{liu2019leaking}                 & \begin{tabular}[c]{@{}l@{}}Period of a\\ victim task\end{tabular}         & \textbf{Multiple} & Known EDF                                                                  \\ \hline  
Babar et al.~\cite{babar2025investigating} & \begin{tabular}[c]{@{}l@{}}Execution mode\\ of a victim task\end{tabular} & Exactly Two       & \begin{tabular}[c]{@{}l@{}}Known \\ fixed-priority\end{tabular}            \\ \hline
\textbf{\sys}                              & \textbf{Sensitive events}                                                 & \textbf{Multiple} & \textbf{\begin{tabular}[c]{@{}l@{}}Not required\end{tabular}} \\ \hline
\end{tabular}
\vspace{1ex}
\caption{Comparison with other schedule-based attacks} \vspace{-7ex}
\label{tab:comparison-related}
\end{table}

Many existing schedule-based side-channel attacks do not capture this threat. They either assume single-mode systems with fixed task parameters~\cite{chen2019novel,chen2020scheduler}, or recover a parameter of one known victim task, such as its period~\cite{liu2019leaking} or execution mode~\cite{babar2025investigating}, and assume the attacker knows how that parameter maps to events of interest. 
In practice, the attacker may know neither which task reflects an event (i.e., which task is the \emph{victim task}) nor the task parameters or scheduling policy, particularly when the system contains closed-source components. 
Defenses against these attacks randomize or reorder the schedule~\cite{yoon2016taskshuffler,chen2018reorder,chen2021indistinguishability}, but they protect identified victim tasks rather than hide the system-wide changes in tasksets and timing parameters that mode changes cause. Table~\ref{tab:comparison-related} contrasts these attacks with our approach. 

Exploiting this multi-mode side channel raises three challenges. First, the attack must work under {\em limited visibility}, where the attacker observes only its own task. This rules out existing mode-inference techniques~\cite{DBLP:journals/rts/VadineanuN22,DBLP:conf/rtas/BrandenburgC0Y25} as they require detailed execution traces of \emph{all} tasks. The mapping from mode changes to events may also be unknown and need not be one-to-one: a single event may span multiple changes, and an individual mode change may not uniquely identify the event. 
Second, the attack must work in the presence of {\em noise}:
it must identify event-related timing patterns despite
scheduling jitter and execution variability unrelated to the events. Finally, the attack must be {\em practical}: it must run with low overhead on resource-constrained devices and transfer to victim hardware that differs from the attacker's profiling devices~\cite{DBLP:conf/ndss/BhasinCHJPS20}. 
 
We present \sys, a \underline{s}chedule‑based \underline{t}iming side‑channel \underline{a}ttack tailored to \underline{m}ulti‑mode \underline{p}latforms. \sys targets the system's modes rather than a single victim task, using one of the two strategies chosen by what the attacker knows about those modes (Sec.~\ref{sec:stamp-attack}). 
When modes are \emph{explicit}, i.e., labeled and with a known mapping to events,
\sys reconstructs the mode trajectory in two steps. It detects mode changes as shifts in windowed timing percentiles, which are robust to transient jitter, and then classifies each segment using timing and core-placement features aligned across devices. When modes are \emph{latent}, \sys skips mode reconstruction and classifies each observation window directly into an event, so that patterns spanning several mode changes identify the event jointly. 
In both cases, \sys normalizes its features against unlabeled measurements
recorded on the target device during a calibration phase before the attack (for
latent modes, also via within-trace ranks), so that models trained on
attacker-owned devices transfer to unseen hardware.

We evaluated \sys on three real-world setups and 6,400 synthetic tasksets. On a home-security system running \zm~\cite{zoneminder}, \sys identifies camera modes, a proxy for occupancy, with 94.7\% accuracy on hardware unseen during training, using under 0.5\% of one core. On ROS~2 with Adaptive Monte Carlo Localization (AMCL) replaying the Oxford RobotCar dataset~\cite{RobotCarDatasetIJRR}, it identifies which of seven route segments a vehicle traversed with 86.0\% accuracy on unseen platforms, compared with 74.6\% reported by a cache side-channel attack on the same dataset without cross-device evaluation~\cite{luo2020stealthy}. On a physical robot running the Timed Elastic Band (TEB) local planner, it identifies six route-and-obstacle scenarios with 88.3\% accuracy. Finally, a representative schedule-randomization defense~\cite{chen2021indistinguishability} lowers \sys's detection recall to as little as 0.53, but only 53.9\% %
of jobs then complete on time in that setting, and up to 14.1\% of tasksets leak only under the defense.

In summary, we make the following contributions:
\wcircled{1} We identify a scheduling side channel created by mode changes in multi-mode CPS that enables an unprivileged task to infer sensitive events from its own execution timing and core assignment, without probing shared memory or caches (Sec.~\ref{sec:attack-overview}). 
\wcircled{2} We design and implement \sys, with two inference strategies for explicit and latent modes that tolerate scheduling noise and transfer across devices (Sec.~\ref{sec:stamp-attack}). 
\wcircled{3} We demonstrate \sys's effectiveness on real-world and synthetic systems, and show that a schedule-randomization defense offers only partial protection, substantially reduces on-time job completions, and can introduce new leakage (Sec.~\ref{sec:case-study-1}--Sec.~\ref{sec:evaluation}).

\section{System and Threat Model}
\label{sec:system-model}

\subsection{Multi-mode system model}
\noindent{\bf Tasks and modes.} We consider a multi-mode real-time system executing a taskset 
$\Gamma = \{\tau_1, \tau_2, \dots, \tau_n\}$ on one or more cores under a deterministic scheduling policy, such as global fixed-priority (FP) or earliest-deadline-first (EDF). The system operates in multiple modes, where each mode $M_m$ executes a subset of active tasks $\Gamma_m \subseteq \Gamma$. 
Each task $\tau_i \in \Gamma_m$ has a mode-specific period, deadline and execution-time distribution. 
A mode change alters the active taskset $\Gamma_m$, the task parameters, or both.

\noindent{\bf Multi-mode execution.} 
Let $W=[0,t_{\max}]$ be an observation window representing the duration of an observed trace, where $t_{\max} > 0$, and $\mathcal{M}$ be the set of system modes. We represent the mode executed at each time by a {\em mode trajectory} function
$
\mu:W\rightarrow\mathcal{M},
$
where $\mu(t)$ is the system mode at relative time $t\in W$. 

\noindent{\bf Sensitive events.} 

Let $V \in \mathcal{S}_V$ be the sensitive-event information
associated with the system execution over the observation window $W$.
The output space $\mathcal{S}_V$ specifies the information targeted
by the attacker and is fixed for a given attack objective.
For example, $V$ may identify the route traversed during $W$
or describe a sequence of sensitive events and their occurrence times.

Let $\mathcal{S}_\mu$ denote the set of feasible mode trajectories
over $W$ and $\Delta(\mathcal{S}_V)$ denote the set of probability distributions over $\mathcal{S}_V$. We model the relationship between mode trajectories and
sensitive events using a stochastic mapping
$\Phi:\mathcal{S}_\mu \to \Delta(\mathcal{S}_V)$,
where $\Phi(\mu)$ is the conditional distribution of $V$ given
the mode trajectory $\mu$.

\noindent{\bf Explicit vs.\ latent modes.} 
We distinguish two cases based on the mode information available to the attacker. A system has {\em explicit} modes if the mode set and labels, as well as the mapping $\Phi$ from mode trajectories to sensitive events, are known to the attacker. Otherwise, the system has {\em latent} modes. For example, a home-security system may expose labeled modes such as \emph{home} and \emph{away}, with transitions corresponding to user presence or absence. In contrast, autonomous-driving and robotic platforms may contain internal modes that are not exposed to the attacker and whose transitions do not directly correspond to sensitive events such as a vehicle's route. In the explicit-mode case, \sys first recovers the mode trajectory and then uses $\Phi$ to infer sensitive events. In the latent-mode case, \sys bypasses mode reconstruction and infers sensitive events directly from observable timing.

\begin{figure*}[t]
    \centering
    \includegraphics[width=1\linewidth, trim={0ex 1.5cm 0ex 0ex}, clip]{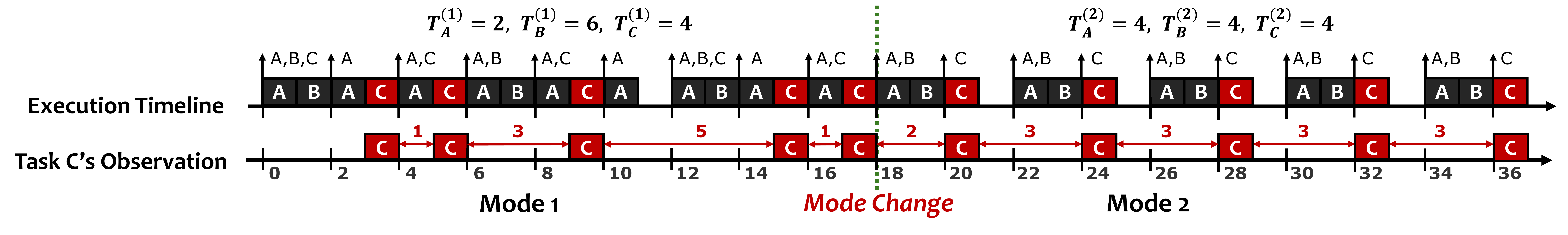}
    \vspace{-3ex}
    \caption{An example of changes in timing patterns due to a mode change, where changed tasks release their new jobs based on new periods. Task C observes a change in its own execution timing pattern due to the mode change.}
    \vspace{-2ex}
    \label{fig:schedule-example}
\end{figure*}

\subsection{Threat model and goal}

\bsection{Observer task}
Following prior side-channel threat models~\cite{ghassami2015capacity,DBLP:conf/sp/JanaS12a,DBLP:conf/ccs/Naghibijouybari18,chen2019novel,chen2020scheduler,DBLP:conf/uss/ZhangSCA23,luo2020stealthy,DBLP:conf/date/WangHPCSH22}, we assume that the attacker can inject code that runs either inside an ordinary, unprivileged task or as a new task; this task serves as the \emph{observer task}. This capability may arise through a compromised software dependency or third-party application~\cite{supplychain-guardian,10.1007/978-3-030-52683-2_2}. 
The observer runs at user level under a trusted operating system and scheduler. 
It records the start and finish times of its own jobs using \texttt{clock\_gettime()} and the corresponding CPU cores using \texttt{sched\_getcpu()}. 
It cannot read another task's memory, internal state, or scheduling statistics; nor can it alter scheduling parameters, raise its priority, or control mode changes. We also exclude direct data theft and reliable denial of service, such as deliberately starving a protected task of CPU time, from the attacker's capabilities.

We assume that the platform's memory management unit (MMU) or memory protection unit (MPU) isolates the observer from privileged kernel data structures holding scheduler state and task parameters~\cite{DBLP:conf/usenix/HardinSPHSK18}. We further assume that debugging and firmware-readout interfaces are disabled or restricted on the victim device~\cite{stm-memprotect}. \sys also needs no system-wide CPU or load statistics, which deployments increasingly restrict: for instance, Android has limited third-party apps' access to them since Android 6~\cite{DBLP:conf/ccs/SpreitzerKGM18}, and since Linux 5.8 a single procfs mount option (\texttt{subset=pid}) hides them from containers and sandboxed programs~\cite{gladkov2020procfs}.  
The observer's own execution timing and core assignments remain available under these restrictions. \sys uses these measurements for local inference and exfiltrates only the resulting low-bandwidth event estimates. If the observer lacks direct network access, it can relay these estimates through existing local communication or covert channels~\cite{hielscher2021systematic,DBLP:conf/uss/ShahM06,DBLP:journals/cybersecurity/CabajZNPM20}.

\noindent{\bf Preparation phase.} 
During an offline preparation phase, the attacker triggers controlled events 
on one or more attacker-owned devices and records the observer's timing traces, which are labeled with the events and used to train the inference models. 
These devices run the same applications as the victim device but may differ in hardware, so the models must transfer across devices. 
Such profiling is common in hardware and physical side-channel  attacks~\cite{luo2020stealthy,DBLP:conf/sp/BasakG22,DBLP:journals/corr/ApthorpeRF17,DBLP:conf/uss/ZhangCGY024,DBLP:conf/uss/HorvathL0B24,DBLP:conf/uss/ZhangSCA23,IoTMosaic,DBLP:conf/acsac/DiptaG22,DBLP:conf/sp/ZhanZLYK22,DBLP:conf/ndss/BhasinCHJPS20}. 

\noindent{\bf Calibration phase.}
Before launching an actual attack, the observer task records its own execution timing and core assignments on the victim device while the victim operates as usual (e.g., the vehicle drives its usual routes). \sys uses these
recordings to estimate device-specific reference statistics that let 
models trained on attacker-owned devices transfer to the victim's
hardware. The recordings are unlabeled. Calibration requires no capabilities beyond those of the attack phase, only additional observation time.

\noindent{\bf Attack phase and attack goal.} 
At runtime, the observer records its own execution timing and core
assignments $\mathbf{X}$ on the victim
device over the observation window $W$. Using models
trained during preparation and the reference values obtained
during calibration, the attacker computes from $\mathbf{X}$ an
estimate $\widehat{V} \in \mathcal{S}_V$ of the sensitive-event information $V$.

\section{Observations and Attack Overview}
\label{sec:attack-overview}

\bsection{Effects of mode changes} 
In multi-mode systems, mode changes may alter task {\em timing behavior} or {\em core-assignment} patterns. To illustrate the effect on timing patterns, Fig.~\ref{fig:schedule-example} shows a fixed-priority task system with two modes and three tasks active in both modes, with priorities $A>B>C$. Each task has execution time 1. In Mode 1, Tasks A and B have periods $2$ and $6$, respectively; in Mode 2, both have period $4$. Task C has period $4$ in both modes. When the system switches from Mode 1 to Mode 2, the changed release patterns of Tasks A and B alter the scheduling slots available to Task C, as shown in the figure. More generally, changes in the active taskset, task periods, deadlines, or execution-time distributions can change the interference experienced by a co-running task and thereby alter its observable timing.

On multicore platforms, mode changes can also change task-to-core  assignment behavior. Changes in the active workload across modes can alter per-core contention and processor availability, which in turn affect scheduler   placement and migration decisions. Consequently, an observer task may be assigned to a different set of cores, exhibit a different distribution of assignments across cores, or migrate among cores at a different rate after a mode change. These changes do not directly reveal the state of other tasks; rather, they provide an additional observable signature of the mode-dependent workload and scheduling behavior.\\[-2ex]

\bsection{\sys Overview} 
Because mode changes alter the observer's timing and core assignments,
the observer records, for each of its jobs, the start time, finish time,
and the cores on which the job starts and finishes. \sys exploits the
resulting changes in features extracted from these measurements to infer
sensitive events.

\emph{1) Preparation phase.}
The attacker first collects labeled traces on the device(s) they own for offline training. 
They run the observer task to record the timing and core assignment sequence $\mathbf{X}$ of its jobs, while simultaneously triggering or recording the events of interest. 
Based on the mode information available to the attacker, \sys selects the corresponding attack strategy and trains the associated inference models.

\emph{2) Calibration phase.} Before the attack, the observer
records measurements on the victim device without performing inference. \sys then aggregates them into device-specific reference statistics
that make features comparable across devices, so that models trained in the
preparation phase transfer to the victim's hardware.

\emph{3) Attack phase.} 
On the victim device, the observer task records its execution timing and core assignment measurements and runs the lightweight inference pipeline to predict events from features extracted from these observations. 
Only compact outputs, such as inferred event labels and timestamps \(\widehat{V}\), are exfiltrated. 
The models and feature extraction are designed to operate within a small CPU budget to minimize scheduling and resource overhead and to avoid triggering usage-based anomaly detectors. \\[-2ex]

\bsection{Attack boundary and limitations}
The attack boundary of \sys is fundamentally determined by the observability and distinguishability of mode-dependent scheduling behavior. 
\sys is effective only when the scheduling behavior associated with sensitive events produces distinguishable timing
or core-assignment patterns at the unprivileged observer.
Two semantically distinct modes may nevertheless produce similar observable patterns and become difficult or impossible to distinguish. 
The attack also relies on a stable stochastic relationship between the mode trajectory $\mu$ and the sensitive-event information $V$. 
Its accuracy thus degrades when the conditional distributions $\Pr(V\mid\mu)$ for different events overlap substantially. \sys further assumes that the feature distributions learned during offline profiling remain sufficiently representative of the victim environment; substantial changes in workload, scheduler configuration, or deployment conditions may reduce accuracy.

\section{The \sys Attack}
\label{sec:stamp-attack}

\subsection{Timestamp collection and preprocessing}
\noindent The observer task records two timestamp sequences and two core assignment sequences during its execution:
\begin{align*}
    t_s = \{t_s(j)\}_{j=1}^{N}, &\quad t_e = \{t_e(j)\}_{j=1}^{N}, \\
\mathsf{cpu}_s = \{\mathsf{cpu}_s(j)\}_{j=1}^{N}, &\quad \mathsf{cpu}_e = \{\mathsf{cpu}_e(j)\}_{j=1}^{N}
\end{align*}
where \(t_s(j)\) and \(t_e(j)\) denote the {\em start} and {\em end} times of its \(j\)-th job, respectively, while \(\mathsf{cpu}_s(j)\) and \(\mathsf{cpu}_e(j)\) denote the cores on which the job starts and finishes, respectively. The observer task can access the system clock and determine its core assignment, but has no privileged visibility into scheduler events such as preemptions or resumptions.
From these timestamps, the observer constructs the measurement sequence
\begin{align*}
&\mathbf{X} = \{\mathbf{x}[j]\}_{j=1}^{N-1}, \quad 
\mathbf{x}[j] = 
\begin{bmatrix}
T_s(j) \\ T_r(j) \\ \mathsf{cpu}_s(j) \\ \mathsf{cpu}_e(j)
\end{bmatrix},
\end{align*}
where 
\begin{align*}
&T_s(j) = t_s(j+1) - t_s(j) \quad \text{(start-to-start interval)},
\\
&T_r(j) = t_e(j) - t_s(j) \quad \quad \text{(start-to-finish interval)}.
\end{align*}

$T_s(j)$ reflects interference between consecutive observer jobs, while $T_r(j)$ captures execution and preemption delays experienced by job $j$. 
Together, these measurements encode scheduling interference from co-running tasks.

\subsection{\sys against explicit-mode systems} \label{subsec:explict-attack}
\subsubsection{Mode change detection} \label{subsubsec:cpd}
\begin{figure}[t]
    \centering
    \includegraphics[width=.8\linewidth, trim={0ex 1mm 0ex 0ex}, clip]{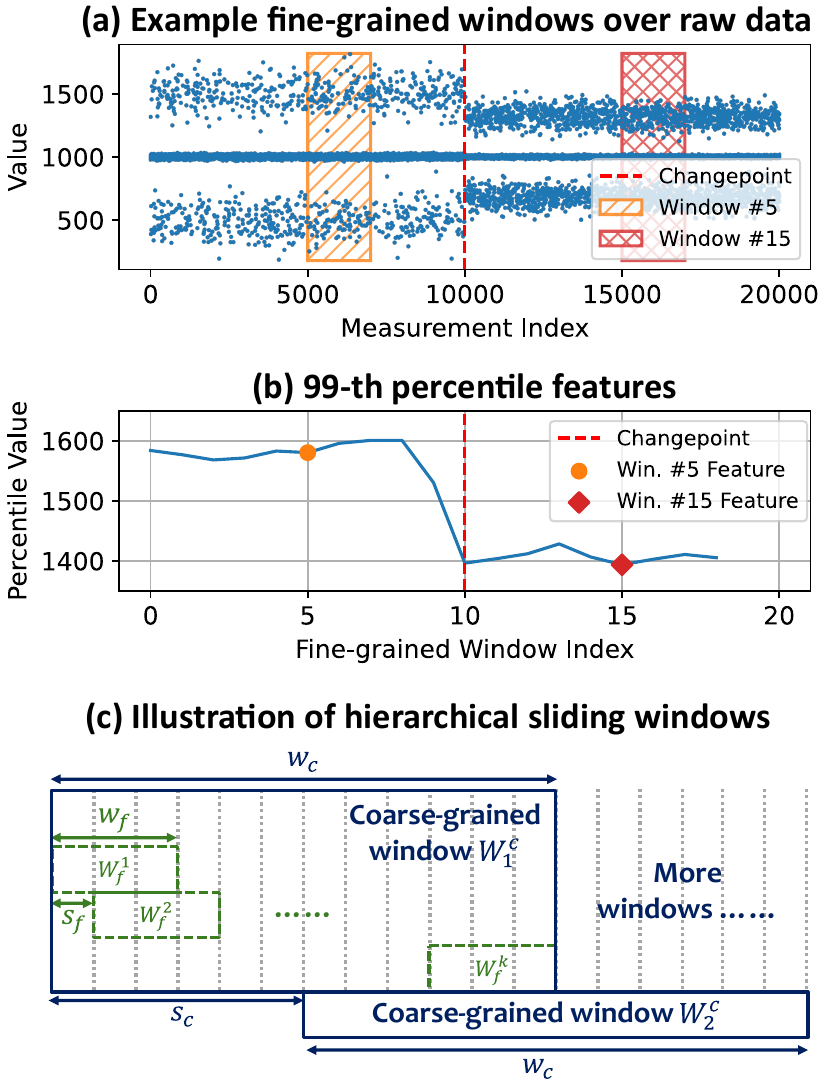}
    \vspace*{-2ex}
    \caption{Sliding windows and percentile feature illustrations. The red vertical dashed line indicates a changepoint.}
    \label{fig:example-mc}
    \vspace*{-3ex}
\end{figure}

\mbox{}\\ 
\bsection{Changepoint detection and its formulation}
As discussed in Sec.~\ref{sec:attack-overview}, in systems with explicit modes, the attacker first determines when each mode change occurs. We formulate this as a changepoint detection (CPD) problem~\cite{DBLP:journals/csda/CelisseMPR18,killick2012optimal,fryzlewicz2014wild,fryzlewicz2007unbalanced} over \(\mathbf{X}\), as illustrated in  Fig.~\ref{fig:example-mc}(a). 
Given a sequence \(\mathbf{X}_{1:N-1}\), CPD estimates a set of ordered changepoints 
\(\mathcal{K} = \{\kappa_1, \dots, \kappa_{c-1}\}\), where \(1 < \kappa_1 < \dots < \kappa_{c-1} < N-1\), which partition $\mathbf{X}$ into $c$ contiguous segments (corresponding to modes):
$\mathbf{X}^{(i)} = \mathbf{X}_{(\kappa_{i-1}+1):\kappa_i}$, where $1\leq i \leq c$.
Let \(\mathcal{C}(\cdot)\) be a segment cost function 
and \(\beta > 0\) be a penalty to avoid over-segmentation. 
We use the standard objective 
\[
\min_{\mathcal{K}}
\Bigg[
  \sum_{i=1}^{c} \mathcal{C}\big(\mathbf{X}_{(\kappa_{i-1}+1):\kappa_i}\big)
  + \beta (c-1)
\Bigg] %
\]
where \(\kappa_0 = 0\) and \(\kappa_c = N-1\) for notational consistency.

A common choice of cost function $\mathcal{C}$ is the within-segment sum of squares:
\(
\mathcal{C}(\mathbf{X}_{a:b}) = \sum_{j=a}^{b} \|\mathbf{x}[j] - \bar{\mathbf{x}}_{a:b}\|_2^2,
\)
which detects mean shifts efficiently.
Further, the attacker cannot assume a known number of changepoints, and the algorithm must operate with low latency. We therefore adapt the Pruned Exact Linear Time (PELT) algorithm~\cite{killick2012optimal}, which supports unknown changepoint counts and achieves linear-time complexity for simple segment costs. 

\bsection{Practical challenges of CPD}
Directly applying PELT to the timing traces $\mathbf{X}$ performs poorly for two reasons. First, its runtime overheads become prohibitive on long traces, especially when using expressive multivariate costs beyond simple mean-shift statistics, for which the linear-time complexity no longer holds. 
Second, timing traces often contain substantial jitter and preemption-induced outliers. Because PELT is sensitive to outliers, such behavior can trigger false changepoints, while aggressive filtering may suppress true ones. 
We further observe that {\em outlier patterns themselves may differ across modes and are important indicators of mode changes}. For 
instance, one mode may produce infrequent but extreme delays, whereas another may produce frequent moderate delays (Fig.~\ref{fig:example-mc}). Since these modes can have similar means and variances, traditional mean-variance cost functions often fail to distinguish them effectively. 
(See Sec.~\ref{subsec:eval-base} for performance results when applying PELT directly to $\mathbf{X}$.)

\bsection{Hierarchical sliding-window detection approach} %
\sys addresses both challenges with two levels of sliding windows:

\textit{Fine-grained feature window.}  
To handle outliers, we introduce a fine-grained sliding window with size $w_f$ and step $s_f$, where $w_f$ and $s_f$ denote the number of measurement elements (jobs), and $s_f$ divides $w_f$.  For each window, we extract features such as variance and selected percentiles
from $\mathbf{X}$ to capture both overall distribution and extreme values. 
Fig.~\ref{fig:example-mc} illustrates two example fine-grained windows over sample raw data (\ref{fig:example-mc}a) and the corresponding 99th percentile features (\ref{fig:example-mc}b). 
Let $\mathbf{X}_k^{\mathsf{fg}}$ denote the data in the $k$-th fine-grained window. Its feature vector is computed as follows: 
\[
\mathbf{z}_k = \phi_{e}(\mathbf{X}_k^{\mathsf{fg}}) =
\Big[
\operatorname{var}(\mathbf{X}_k^{\mathsf{fg}}),\ 
\operatorname{perc}_p(\mathbf{X}_k^{\mathsf{fg}})_{p\in\mathcal{P}}
\Big],
\]
where $\mathcal{P}$ is a set of application-dependent percentiles. 
These features summarize both overall variability and tail behavior, so CPD responds to sustained changes in delay patterns rather than to isolated outliers. CPD is applied to the feature sequence $\{\mathbf{z}_k\}$ rather than directly to $\mathbf{X}$.

\textit{Coarse-grained window.}  
To bound runtime and memory usage, \sys maintains a coarse-grained window with size $w_c$ and step $s_c$, where $w_f\mid s_c$ and $s_c\mid w_c$, as illustrated in  Fig.~\ref{fig:example-mc}(c). CPD is invoked only once per coarse-grained window, i.e., after the extracted features span $w_c$ jobs. After each invocation, the oldest features corresponding to $s_c$ oldest jobs are discarded.
This design bounds the working set size and enables continuous online operation with predictable overhead. The parameters $(w_c,s_c,w_f,s_f,\beta)$ are tuned offline on profiling data to balance latency and detection accuracy (cf. Sec.~\ref{subsec:eval-base}).

\noindent{\bf Mapping changepoints to occurrence times.} 
Suppose the $i$-th changepoint is detected in fine-grained window $k$ of coarse-grained window $l$. Then, its observer-job index is $\kappa_i = (l-1) \cdot s_c + (k-1)\cdot s_f$, and its occurrence time is estimated as $t_s(\kappa_i)$.

\subsubsection{Event reconstruction}
\label{subsec:event-reconstruction}\mbox{}\\
\bsection{Feature extraction from detected segments}
Let
\(\mathcal{K}=\{\kappa_1,\dots,\kappa_{c-1}\}\)
be the detected changepoints and
$\mathbf{X}^{(i)} = \mathbf{X}_{(\kappa_{i-1}+1):\kappa_i}$ $(1 \leq i \leq c)$
be the resulting mode segments. For each detected segment, \sys applies a
feature map directly to all of its measurements:
$\mathbf{f}_i = \phi_{\mathsf{cls}} ( \mathbf{X}^{(i)})$.
The feature map summarizes the timing behavior observed over the
entire segment using the following features:
\vspace*{-.5ex}
\begin{itemize}[leftmargin=*]
    \item \emph{Delay frequency} describes how often the start-to-start
    intervals between consecutive observer jobs exceed its period as a
    consequence of resource contention. Delay frequencies at different
    thresholds relative to the period capture both frequent small
    disturbances and less frequent large disturbances.

    \item \emph{Catch-up frequency} describes how often the observer starts
    its next job less than one period after the previous one. Such closely
    spaced starts occur when the observer catches up after an earlier delay
    and therefore provide another indication of system load.

    \item \emph{Delay magnitude and burstiness} describe how much the
    start-to-start intervals exceed the period and whether this excess is
    spread across many jobs or concentrated in a few large excursions.

    \item \emph{Execution stretch} compares each job's start-to-finish time
    with the observer's typical start-to-finish time $\hat{e}$ on the victim
    device, measured as the median over the calibration phase (Sec.~\ref{sec:attack-overview}). Jobs
    that take longer than $\hat{e}$ were delayed while executing, e.g., by
    preemption, so execution stretch captures how often and by how much the
    observer is preempted.

    \item \emph{Core-assignment behavior} summarizes how the observer is
    distributed across cores, including whether execution is concentrated on
    one core and how frequently the observer migrates between cores.
\end{itemize}
Together, these features capture how frequently interference occurs, how
severe it is, how it is distributed over time, and how it affects the
observer's execution and core assignment.

\bsection{Cross-device feature alignment}
Feature values can differ across devices. 
\sys therefore applies $\phi_{\mathsf{cls}}$ to fixed-length windows of
the calibration phase, takes a low and a high percentile (e.g., 2nd and 98th)
of each feature as
reference values, and maps this range to a common scale $[0,1]$, yielding
$\widetilde{\mathbf{f}_i}$. Percentiles, unlike the minimum and maximum, keep
the reference values robust to outliers. If the attacker prepares on
hardware matching the victim's, calibration is skipped and these
reference values, together with $\hat{e}$, are computed from the
preparation traces instead.

\bsection{Mode classification and event reconstruction}
During preparation, \sys pairs each aligned segment-level feature vector
\(\widetilde{\mathbf{f}}_i\) with the known mode in which the segment was
collected. These labeled segment vectors are used to train a classifier
$\psi: \widetilde{\mathbf{f}}_i \mapsto \mathbb{R}^{|\mathcal{M}|}$,
which predicts a probability for each mode in \(\mathcal{M}\). During the attack,
the predicted mode is the mode with the highest probability:
\[
\widehat{M}_i
=
\arg\max_{m\in\mathcal{M}}
\left[
\psi^\star
\left(
\widetilde{\mathbf{f}}_i
\right)
\right]_m,
\]
where \(\psi^\star\) denotes the trained classifier.

Applying the classifier to the \(c\) detected segments produces the
ordered mode trajectory
\(\widehat{M}_1,\dots,\widehat{M}_c\). Together with the changepoints,
this defines the reconstructed mode trajectory
\(\widehat{\mu}\): segment \(\mathbf{X}^{(i)}\) is assigned mode
\(\widehat{M}_i\), and the transition from \(\widehat{M}_i\) to
\(\widehat{M}_{i+1}\) occurs at the estimated time
\(t_s(\kappa_i)\). 
Using the known mapping $\Phi$, \sys infers the
sensitive events from the reconstructed mode trajectory as $\widehat{V} = \arg\max_{V} \Phi(\widehat{\mu})(V)$.

\subsection{\sys against latent-mode systems} \label{subsec:latent-attack}
In latent-mode systems, an event's timing signature may span several
mode changes. Two routes, for example, may produce
similar interference bursts that differ only in {\em when} they 
occur. \sys therefore classifies each complete observation window 
as a whole. Using event-labeled observer traces, it
learns to infer the sensitive-event information $V$ from the window's 
timing pattern.

\bsection{Percentile-based timing representation}
For latent-mode systems, \sys uses different features for the timing components compared to the explicit-mode attack.\footnote{The ablation in Sec.~\ref{subsec:eval-amcl} (Fig.~\ref{fig:ablation-amcl}) shows that the pipeline using explicit-mode features reach only 61.3\% on route prediction, versus 86.0\% for this representation.} Consider:
\[
\widetilde{\mathbf{X}}
=
\{\widetilde{\mathbf{x}}[j]\}_{j=1}^{N-1},
\qquad
\widetilde{\mathbf{x}}[j]
=
\begin{bmatrix}
T_s(j)\\
T_r(j)
\end{bmatrix},
\]
where $T_s(j)$ and $T_r(j)$ denote the start-to-start and start-to-finish intervals, respectively. 

Absolute timing values can vary substantially across devices and
configurations. \sys therefore expresses timing values
by their percentile positions using two reference distributions:
measurements within the current observation window and within unlabeled
measurements from the calibration phase. Both representations are
computed separately for $T_s$ and $T_r$.

Within the observation window, \sys computes percentile
ranks and identifies measurements above selected thresholds,
such as the 90th, 95th, and 98th percentiles. These quantities
describe where relatively large delays occur within the trace.

\bsection{Device normalization}
Ranks within the observation window are relative to the trace itself
and thus cannot tell whether one trace has larger delays overall than
another. \sys therefore also computes each measurement's
\emph{device percentile}, i.e., its percentile within the timing
distribution estimated over the calibration phase rather than the
current observation window, which is too short and too event-specific to
serve as a device reference. Device percentiles place all traces from a
device on a common scale.

\sys additionally calibrates window-level magnitude
statistics, such as upper percentiles and the fraction of
measurements exceeding one or two observer periods, using the
median and interquartile range of the same statistics over
fixed-length windows of the calibration measurements.

\bsection{Features}
For each completed observation window, \sys applies a sliding window of size $w$ and step $s$, where $s$ divides $w$. For each window, it computes:
(i) {\em mean rank}, the mean percentile rank of the measurements in the window;
(ii) {\em tail rank}, the fraction of measurements above each selected percentile threshold; and
(iii) {\em device-percentile features}, the mean device percentile and the fraction of measurements above or below fixed device percentiles. Unlike tail rank, whose thresholds come from the observation window itself, these capture how many delays are large for the device.
Let $\mathbf{z}_i$ denote the features from the $i$-th window. As the
window advances through the trace, the sequence
$\mathbf{F}^{\mathsf{win}} = [\mathbf{z}_1,\mathbf{z}_2,\dots]$
describes how the timing behavior evolves during the observed execution window. For
example, it captures whether large start-to-start intervals occur primarily
near the beginning or end of the trace, whether start-to-finish delays appear
in several distinct phases, and whether large timing values are concentrated
in a short interval or distributed throughout the trace.

Over the entire observation window, \sys additionally computes: (i) how clustered the largest measurements are and where they occur; (ii) summary statistics of the device percentiles, such as the fraction of measurements above or below fixed device percentiles; and (iii) the device-relative magnitude statistics. It also records the duration of the observation window. \sys concatenates $\mathbf{F}^{\mathsf{win}}$ and these features into the feature vector $\mathbf{F}^{\mathsf{lat}}$.

\bsection{Event classification}
Using labeled preparation traces
$(\mathbf{F}^{\mathsf{lat}},V)$, \sys trains a classifier
$\psi_{\mathsf{lat}}:
\mathbf{F}^{\mathsf{lat}}
\mapsto V$, which combines several classifiers, each trained on a subset of $\mathbf{F}^{\mathsf{lat}}$ (e.g., only the window sequence $\mathbf{F}^{\mathsf{win}}$). For each candidate event label, \sys sums the log-probabilities output by the classifiers, each multiplied by a non-negative weight selected on held-out training traces, and predicts the label with the highest score.
During the attack phase, the observer continuously collects timing measurements. Once an observation window is complete, \sys computes the percentile ranks and device percentiles, extracts the features, and infers the sensitive-event information:
$\widehat{V}= \psi_{\mathsf{lat}}^\star \left(\mathbf{F}^{\mathsf{lat}}\right)$.

\section{Case Study 1: Home Security System}\label{sec:case-study-1}
To demonstrate the utility of \sys, we next present evaluation results using two real-world case studies and a range of synthetic workloads. 
We first evaluate \sys on a smart home security system to demonstrate how schedule-based timing leakage can reveal user activity in an explicit-mode system.

\noindent{\bf Default setup.}
Our default setup consists of two {\em Reolink E1 Pro} cameras~\cite{reolink-e1-pro} connected to a 
RaspberryPi 4B (64-bit Cortex-A72 CPU at 1.8GHz and 4 GB RAM) 
monitor node running \zm~\cite{zoneminder} via a {\em Linksys E5400} router. 
As training and testing use the same device in this default setup, \sys does not use the calibration phase here.
We configured \zm as a real-time Linux service with the highest priority (via the $\mathsf{chrt}$ command) under SCHED\_FIFO, and isolated a core via \texttt{isolcpus}. 
Each camera operates in either $\mathsf{None}$ mode, where frames are ignored, or $\mathsf{Modect}$ mode, where frames are decoded and analyzed for motion detection periodically. The system therefore has three modes depending on whether zero, one, or both cameras are active. We triggered randomized mode changes every 30--120\,s. The attacker's goal is to detect mode changes and infer the corresponding modes.

We injected an observer task with a 5\,ms period into the \zm service as a separate process. We used sliding-window sizes $w_c=20000$ and $w_f=100$, with step sizes equal to half of the window sizes. CPD used the $\mathsf{ruptures}$~\cite{ruptures} library, and event reconstruction used an ExtraTrees~\cite{DBLP:journals/ml/GeurtsEW06} classifier with 300 trees.

\begin{figure}
    \centering
    \includegraphics[width=0.8\linewidth, trim={0ex 1.5ex 0ex 1.5ex}, clip]{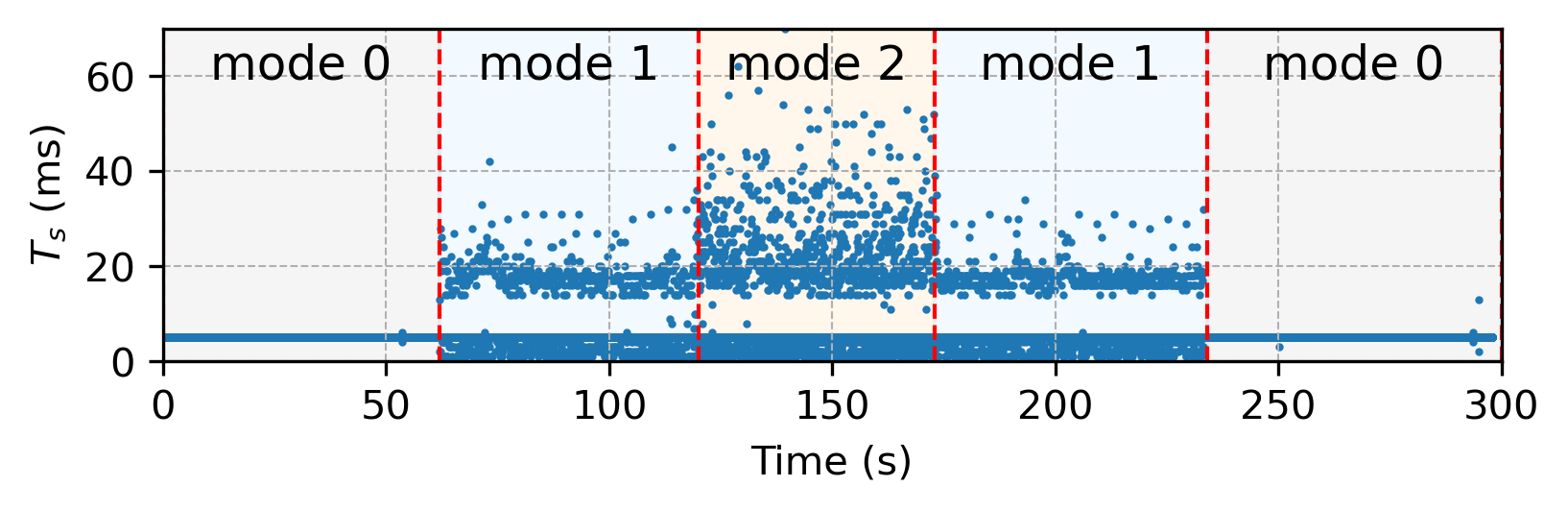}
    \vspace*{-3ex}
    \caption{Time series of $T_s$ values observed under \zm across mode changes.}
    \label{fig:zm-trace}
    \vspace*{-2ex}
\end{figure}

\bsection{Preparation phase}
We collected 30 minutes of observer traces under randomized mode changes to tune the CPD parameters and train the mode classifier. Fig.~\ref{fig:zm-trace} shows an example of the observed $T_s$ values, where different modes produce distinct timing patterns.

\vspace{.25ex}
\bsection{Attack overhead}
Most runtime overhead comes from changepoint detection and mode classification. Under single-core configuration, CPD and classification require 95.94\,ms and 43.82\,ms on average every 50\,s (corresponding to a step size of 10000 jobs). This corresponds to 0.28\% CPU utilization in total. Under a configuration with two cores reserved for \zm, overhead increases to 146.66\,ms and 59.36\,ms due to the additional CPU-core features. Overall, \sys operates in real time on the embedded platform with negligible overhead.

\begin{figure}
    \centering
    \includegraphics[width=0.8\linewidth, trim={0ex 1ex 0ex 1ex}, clip]{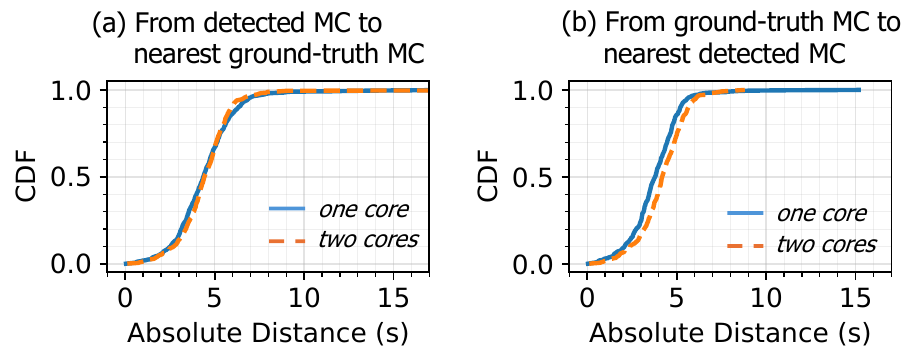}
    \vspace*{-3ex}
    \caption{Distances from detected MCs to ground truths.}
    \label{fig:zm-distance-cdf}
    \vspace*{-3ex}
\end{figure}
\vspace{.25ex}
\bsection{Accuracy} 
Fig.~\ref{fig:zm-distance-cdf} shows the CDFs of the distances between detected and ground-truth mode changes. The results show that over 99\% of ground-truth mode changes are detected within 7.84\,s, and over 99\% of detected changepoints correspond to a true mode change within 9.33\,s. \sys further achieves mode-classification accuracy of 100.0\%.
Using the detected changepoints and inferred modes, \sys reconstructs the mode trajectory for all modes with high accuracy. During a 24-hour experiment, the reconstructed trajectory matches the ground truth for 97.6\% of the time.

\bsection{Robustness against system configurations}
We evaluated \sys's robustness under software and system-configuration changes. We trained the models under the $\mathsf{ondemand}$ DVFS setting and \zm v1.36. We then upgraded \zm to v1.38, switched DVFS to $\mathsf{performance}$, and ran \sys with the trained models continuously for 30 days. Mode trajectory reconstruction accuracy changed by less than 0.2\%, indicating strong robustness across long-term deployment and configuration changes.

\bsection{Transferability across devices and observer configurations}
In practice, an attacker may not have the same hardware as the victim. In addition to the Raspberry Pi 4B used in our default setup, we evaluate \sys on a desktop equipped with an Intel Core i7-4790 processor and 32 GB of memory, and a TinkerBoard~S equipped with a 32-bit Rockchip RK3288 processor and 2 GB of memory. We run each device with one and two isolated CPU cores, producing six hardware scenarios in total. We further vary the observer's period over \(\{5,20,50,100\}\) ms and its utilization over \(\{0,0.2,0.4\}\), producing 12 observer configurations and 72 runs across the six hardware scenarios. We use the same feature and model configuration for all observer configurations and train a separate classifier for each period-utilization pair. We evaluate transferability using leave-one-device-out validation: for each pair and each fold, we train on both core configurations of two devices and test on both configurations of the remaining, unseen device. For each run, the first half of the trace serves as the calibration phase; the second half is used for training or testing, depending on the fold. 
Across all 12 observer configurations, \sys achieves an average cross-device mode-classification accuracy of 94.7\% ($\pm$0.7\%, 95\% Wilson score interval; likewise for all classification accuracies hereafter). This demonstrates that \sys remains effective across different observer configurations and hardware platforms.

\begin{figure}[t]
    \centering
    \includegraphics[width=0.85\linewidth, trim={0ex 2ex 0ex 0ex}, clip]{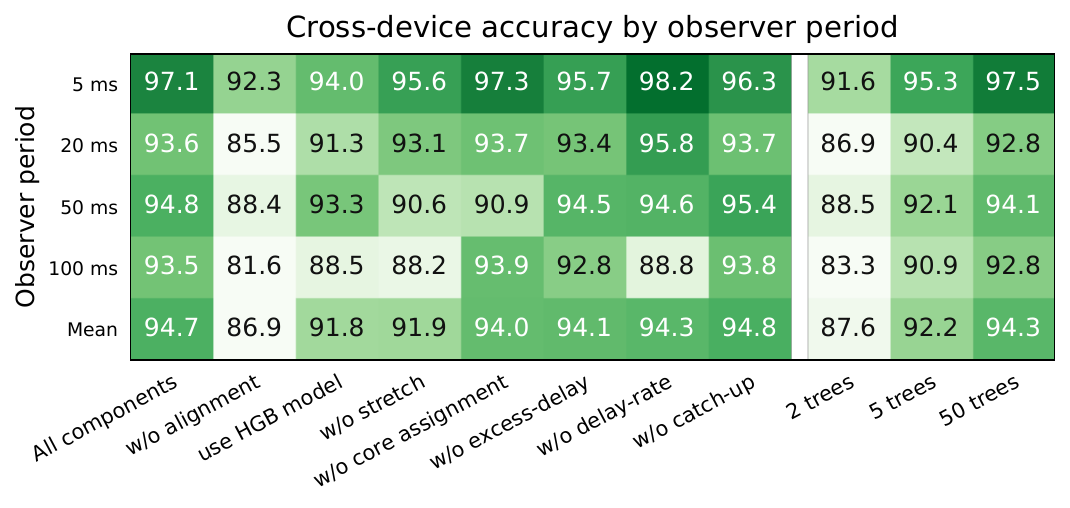}
    \includegraphics[width=0.85\linewidth, trim={0ex 2ex 0ex 0ex}, clip]{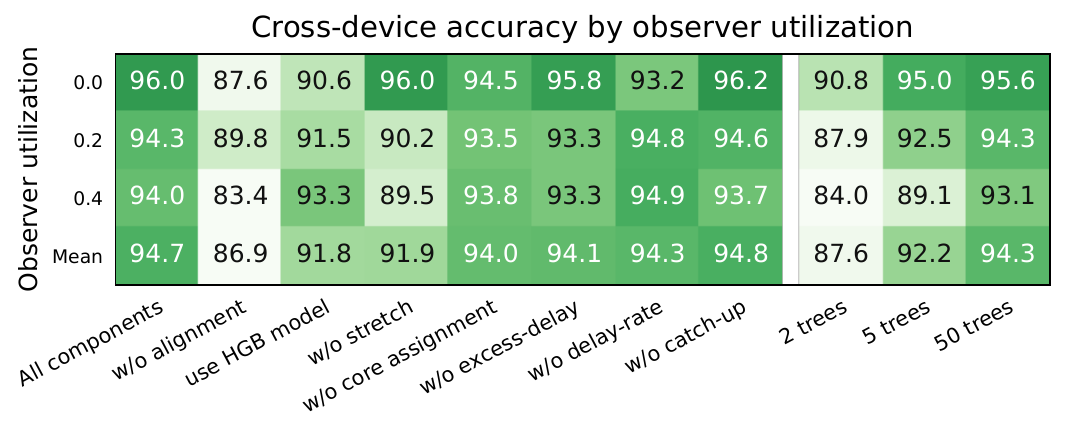}
    \vspace*{-2.5ex}
    \caption{Cross-device mode inference accuracy across different observer configurations, features, and parameters.}
    \label{fig:ablation-zm}
    \vspace*{-3ex}
\end{figure}

\bsection{Ablation and sensitivity study}
Figure~\ref{fig:ablation-zm} evaluates cross-device transfer while holding the
observer's period and utilization fixed between training and test, removing one
component at a time and varying the size of the forest; all variants use the
same calibration phase and test segments as above.
Per-scenario alignment is the dominant contributor: without it, accuracy drops
by 7.8\%, showing that compensating for each device--core configuration's
offset and gain is essential for transfer. Replacing
ExtraTrees~\cite{DBLP:journals/ml/GeurtsEW06} with the HGB
classifier~\cite{DBLP:conf/nips/KeMFWCMYL17,HGB} and removing the
execution-stretch features cost 3.0\% and 2.9\%, respectively, followed by the
core-assignment features (0.8\%).
The stretch and delay-rate features complement each other across observer
utilizations. Execution stretch captures how often the observer is preempted
while executing, which is not observable when the observer's utilization is 0:
removing it has no effect there, but costs 4.1\% and 4.4\% at utilizations 0.2
and 0.4. Conversely, removing the delay-rate features costs 2.7\% at
utilization 0, where delay rate is the main remaining indicator of
interference, but slightly improves accuracy at higher utilizations, for a
0.4\% loss overall. The catch-up features have no noticeable effect on average.
The forest also shrinks at a cost: 50 trees instead of 300 cost 0.4\%, while
5 trees cost 2.5\% and 2 trees cost 7.2\%.

\section{Case Study 2: Vehicle Route Prediction} \label{subsec:eval-amcl}
\subsection{AMCL with the Oxford RobotCar dataset}
\noindent{\bf Platform.} We first consider Adaptive Monte Carlo Localization (AMCL), a widely used localization algorithm in autonomous-driving systems~\cite{chung2021improved,he2023enhanced}, including commercial platforms such as Baidu Apollo~\cite{DBLP:conf/icra/WanYCLZWS18}.
We used real-time Linux and the ROS~2 platform~\cite{ros2-kilted},
which is a standard combination for autonomous and cyber-physical systems used by companies such as Boeing, BMW, Volvo, and Tier~IV (Autoware)~\cite{ros-industrial,DBLP:conf/iccps/KatoTMMHKMAFA18}. Our testbed runs ROS~2 on a real-time Linux kernel (with the $\mathtt{PREEMPT\_RT}$ patch) under fixed-priority preemptive scheduling ($\mathtt{SCHED\_FIFO}$) with CPU isolation, so the tasks under attack are scheduled under the same model assumed in Sec.~\ref{sec:system-model}.
We use the nav2 implementation of AMCL, whose inter-job arrival time depends on the vehicle's speed and changes in the environment.
This case study evaluates whether the resulting observer timing patterns reveal the vehicle's route when mode labels and the mapping $\Phi$ are unavailable, demonstrating a latent-mode system.
We also generated a map using SLAM Toolbox~\cite{DBLP:journals/jossw/MacenskiJ21} in ROS~2 as a required input for AMCL.

\bsection{Dataset}
We used real-world vehicle traces from the Oxford RobotCar dataset~\cite{RobotCarDatasetIJRR}, which contains multiple traversals of the same route.
Following~\cite{luo2020stealthy}, we partitioned the route into seven segments (Fig.~\ref{fig:oxford} in Appendix~\ref{sec:appendix-amcl}) and treated them as distinct routes for prediction.\footnote{As in~\cite{luo2020stealthy}, we assume that each observation window $W$ covers exactly one route. This is realistic because AMCL updates its particle filter only after the vehicle moves beyond a distance or rotation threshold (\texttt{update\_min\_d}, \texttt{update\_min\_a}), so the observer experiences little interference from AMCL while the vehicle is stopped. When a route starts and ends with the vehicle
stopped, as in delivery trips, its start and end appear as the onset and
cessation of interference in the observer's own start-to-start and
start-to-finish intervals, without requiring any privileged information.}
In total, we extracted 126 usable traces with sensor data of good quality spanning diverse traffic and weather conditions.
This setup reflects real-world autonomous-driving and delivery systems, such as those deployed by Uber Eats~\cite{avride}, Starship~\cite{starship-delivery}, and Cainiao~\cite{meituan-delivery}, which repeatedly traverse the same neighborhood routes.
Leakage of such route information may reveal sensitive operational and customer information, creating privacy risks. 

\bsection{Attack setup}
We added the observer to the AMCL package as a separate executable, so that the
package's launch file starts it as its own process alongside the AMCL node.
The observer runs in its own address space and uses no ROS~2
communication. We assume that ROS~2 access control (SROS2) prevents it from
subscribing to localization topics. Since AMCL is single-threaded,
all processes, including the observer, share one reserved core. The observer
runs under \texttt{SCHED\_FIFO} at a priority below AMCL's.
We used the platforms summarized in
Table~\ref{tbl:amcl-platforms} to collect the observer traces.

\begin{table}[t]
\small
\centering
\scalebox{0.9}{
\begin{tabular}{|c|c|c|c|}
\hline
\textbf{Name}    & \textbf{CPU}       & \textbf{Frequency} & \textbf{Memory} \\ \hline
\textbf{Desktop} & Intel Core i7-4790 & 3.6 GHz            & 32 GB           \\ \hline %
\textbf{c6320}   & Intel E5-2683 v3   & 2.0 GHz            & 256 GB          \\ \hline
\textbf{c6525}   & AMD EPYC Rome      & 2.8 GHz            & 128 GB          \\ \hline
\textbf{m510}    & Intel Xeon D-1548  & 2.0 GHz            & 64 GB           \\ \hline
\textbf{xl170}   & Intel E5-2640v4    & 2.4 GHz            & 64 GB           \\ \hline
\end{tabular}
}
\caption{Platforms used for evaluations of the AMCL case study, including one desktop and four servers on CloudLab~\cite{cloudlab}.}
\label{tbl:amcl-platforms} \vspace*{-4ex}
\end{table}

\begin{figure}[t]
    \centering
    \includegraphics[width=.8\linewidth, trim={0ex 1ex 0ex 1ex}, clip]{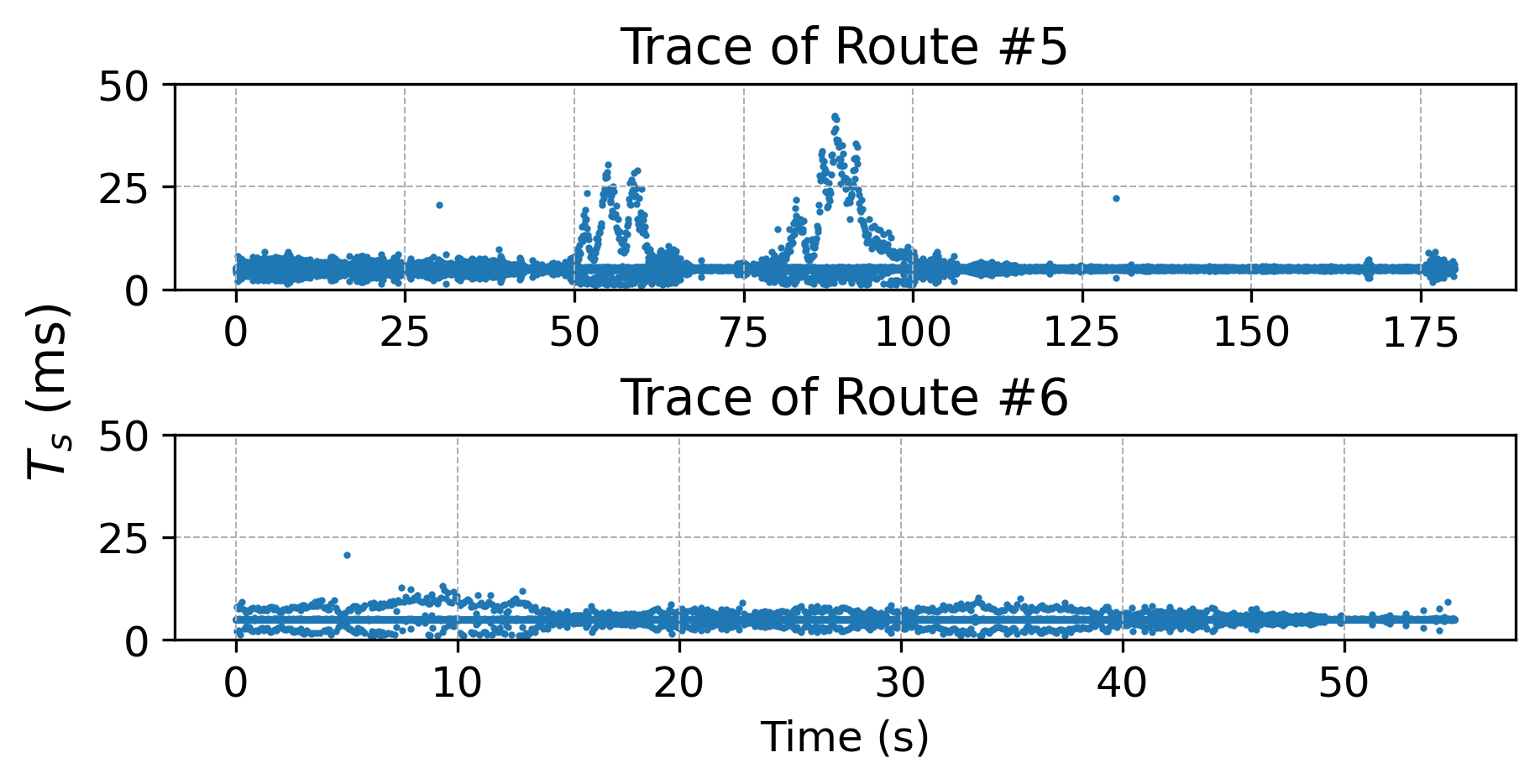} \vspace*{-3ex}
    \caption{Collected traces under two different vehicle routes.}
    \label{fig:ros-trace}  \vspace*{-3ex}
\end{figure}

\noindent {\bf Training.}
 We evaluated 12 configurations of the observer, with its period in \(\{5,20,50,100\}\) ms and its utilization over \(\{0,0.2,0.4\}\). We collected 126 labeled observer traces for each configuration and each device, resulting in $12\times5\times126=7560$ traces in total. Fig.~\ref{fig:ros-trace} shows an example of traces collected with an observer with a 5\,ms period and 0 utilization on the desktop.
To avoid data leakage, every test trace is unseen in both device and trace: for each configuration, we split the 126 traces into ten route-stratified folds and, for each held-out platform and fold, train on the other four platforms using only the other nine folds, so a trace for testing is never seen on any platform. 
As the calibration phase, the attacker uses unlabeled observer data collected
on the test platform under the same configuration, excluding the test trace, for device normalization (Sec.~\ref{subsec:latent-attack}).
\sys combines multiple classifiers: ExtraTrees~\cite{DBLP:journals/ml/GeurtsEW06}, gradient-boosted tree classifier (HGB)~\cite{DBLP:conf/nips/KeMFWCMYL17,HGB}, and a compact 1-D convolutional neural network (CNN). Each classifier may have a different weight in each configuration, chosen on held-out traces within the training data. More details on the classifier design are described in Appendix~\ref{sec:appendix-amcl}.

\noindent {\bf Accuracy.}
Tested on an unseen platform and unseen traces, \sys identifies which of
the seven route segments a vehicle traversed with an accuracy of 86.0\% ($\pm$0.8\%). Accuracy peaks at 88.6\% ($\pm$1.5\%) for the smallest observer period (5\,ms) and remains above 83.3\% ($\pm$1.8\%) for others, and observer utilization has little effect. 

\begin{figure}[t]
    \centering
    \includegraphics[width=0.8\linewidth, trim={0ex 2ex 0ex 0ex}, clip]{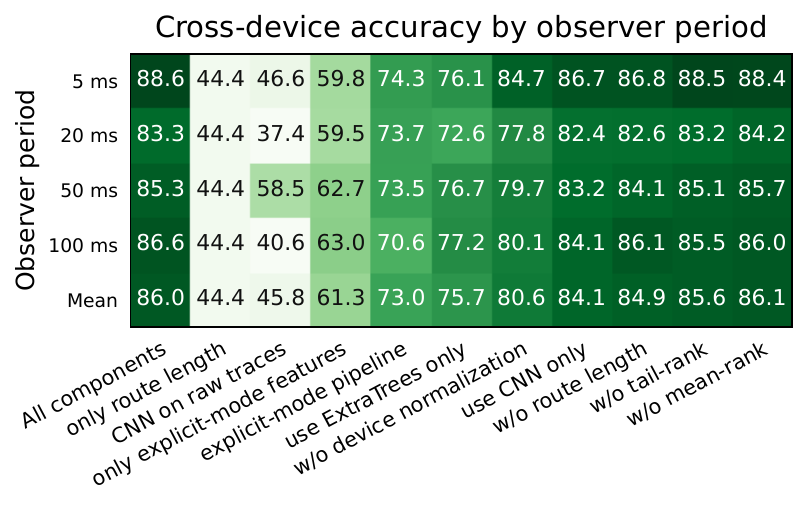}
    \includegraphics[width=0.8\linewidth, trim={0ex 2ex 0ex 0ex}, clip]{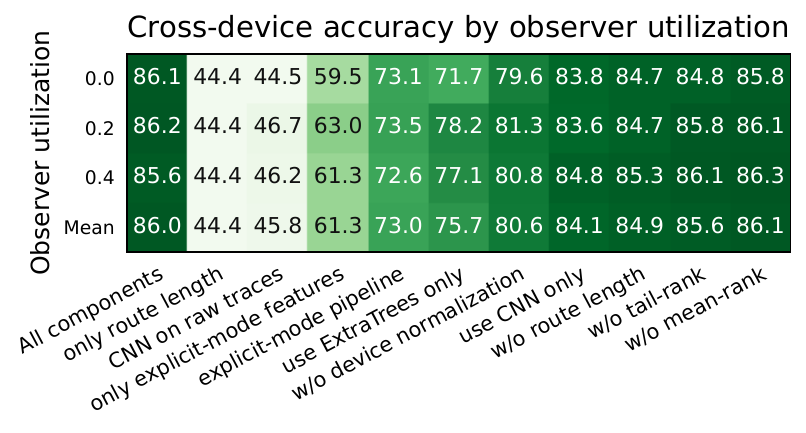} \vspace*{-3ex}
    \caption{Cross-device accuracy of route prediction across observer periods and utilizations with different ablations.}
    \label{fig:ablation-amcl}  \vspace*{-3ex}
\end{figure}

\noindent {\bf Ablation study.} 
Fig.~\ref{fig:ablation-amcl} shows the effect of removing or replacing
one component at a time, relative to the full model (86.0\%). Three factors matter most. First, the explicit-mode attack (Sec.~\ref{subsec:explict-attack}) does not carry over to routes: it reaches 61.3\% ($\pm$1.1\%)  with its features alone and 73.0\% ($\pm$1.0\%)  with its complete pipeline. Its features reduce each window to a load level, which separates modes, whereas a route is identified by how the load varies within the trace. Second, this variation must be resolved at a fine time scale: without the CNN, which sees the trace in 1-s windows, accuracy drops to 75.7\% ($\pm$1.0\%), likely because short load changes are averaged out in the coarser windows of the tree-based classifiers. Third, without device normalization, accuracy drops to 80.6\% ($\pm$0.9\%):  ranks show where delays are large within a trace but not how large they are, so routes with similar timing patterns but different load intensities become hard to distinguish. Finally, two strawman baselines, one using only route duration and one applying a CNN to the raw measurements, reach only 44.4\% ($\pm$1.1\%) and 45.8\% ($\pm$1.1\%) accuracy, respectively.

\noindent {\bf Comparison with other attacks.}
We compared against~\cite{luo2020stealthy}, 
which infers routes on the same workload and dataset by probing a shared cache and reports 74.6\% accuracy without evaluating transfer to unseen devices.
In contrast, \sys needs only the observer's own timing, reaching 86.0\%
($\pm$0.8\%) under a stricter protocol in which the platform is unseen during training. 

\noindent {\bf Prediction latency.}
On a Raspberry Pi 4B, the average prediction latency ranges from 34.4\,ms to 82.4\,ms across observer periods; it is highest at the shortest observer period (5\,ms), where the observer collects the most measurements. Since \sys predicts once per route and routes last 40--200\,s, this amounts to 0.02--0.2\%
of one CPU core.  Removing the CNN halves the latency but lowers accuracy by 10.3 percentage points (to 75.7\%).

\vspace{-1ex}
\subsection{Path planning on a real robot}
\bsection{Platform}
We evaluated \sys on a Yahboom MicroROS Pi5 robot~\cite{yahboom-robot} running the TEB collision-avoiding path planning algorithm~\cite{DBLP:conf/robotik/RosmannFWHB12}, whose execution behavior varies with the surrounding environment and planned path. 
The robot uses a Yahboom MS200 LiDAR to provide environmental inputs to TEB and an \textsf{rf2o} module~\cite{DBLP:conf/icra/JaimezMG16} for odometry.  
All software components are ROS2 modules running on a Raspberry Pi 5 with an Arm Cortex-A76 CPU at 2.4~GHz and 8~GB memory. As in the AMCL case study, we injected the observer as a ROS2 module. The observer runs with a 5\,ms period at negligible CPU utilization. 

\bsection{Trace collection}
We considered three different routes (straight, right-turn, and u-turn), each with and without an obstacle, yielding six scenarios. We collected 10 traces per scenario. 
In each run, the robot traverses the assigned route once; in obstacle scenarios, a box is placed along the route, forcing TEB to plan a detour around it. The observer shares a core with the planner, runs at a lower real-time priority, and records its wake-up intervals from the moment the path is dispatched until the robot reaches its goal. 

\bsection{Training}
We used 10-fold cross-validation, where each fold holds out the $i$-th repetition from all 6 scenarios for testing and uses the remaining 9 repetitions (54 traces) for training. Rotating $i$ over all 10 repetitions yields 10 disjoint test sets that together cover the entire dataset. Each fold is balanced, with one trace per scenario in the test set and nine per scenario in the training set. Because each repetition corresponds to a separate run of the system, every test trace comes from a run not used for training.

\bsection{Results}
\sys identifies the six scenarios with 88.3\% accuracy (with 95\% Wilson confidence interval  [77.8\%, 94.2\%]).
The route is correctly identified in 90.00\% ([79.9\%, 95.3\%]) of traces, while obstacle presence is correctly identified in 96.67\% ([88.6\%, 99.1\%]) of traces. These results demonstrate that \sys generalizes to a real-world robotic system.

\section{Evaluation with Synthetic Workloads}\label{sec:evaluation}
In this section, we evaluate the observability of mode changes,  the effectiveness of our changepoint detection approach (Sec.~\ref{subsubsec:cpd}), and the effects of a randomization-based rugdefense~\cite{chen2021indistinguishability}. 

\subsection{Setup}
\bsection{Taskset generation} We generated periodic tasksets with implicit deadlines, drawing periods
either uniformly from $\{1,2,5,10,20,50,100,200,500,1000\}$~ms, common
in automotive systems~\cite{kramer2015real}, or log-uniformly from
1--1000~ms~\cite{DBLP:conf/rtas/BrandenburgC0Y25}. Tasks ran on 1--4 cores. We varied the number of tasks in
$\{5,10,20,50\}$ and the mean utilization per core in
$\{0.2,0.4,0.6,0.8\}$, using Dirichlet-Rescale~\cite{DBLP:conf/rtss/GriffinBD20}
to split utilization among tasks as in~\cite{DBLP:conf/rtns/CaiPT24}.
Each task's execution time follows either $N(\mu_i,(0.2\mu_i)^2)$ or an
equal mixture of $N(0.8\mu_i,(0.08\mu_i)^2)$ and $N(1.2\mu_i,(0.12\mu_i)^2)$,
chosen at random. We generated 50 tasksets per configuration (6400 in
total) and picked one task in each at random as the observer.

\bsection{Mode changes}
Each taskset starts in an initial mode, then switches to a new mode by 
(1) activating or deactivating
some tasks, (2) assigning new periods to some tasks, or (3) changing
some tasks' execution-time means or distributions.
The observer stays active in both modes; otherwise
detection would be trivial or impossible. 
We varied the fraction of affected tasks during a mode change in $\{20,40,60,80\}\%$. We discarded tasksets that
become unschedulable, and simulated each taskset for 600~s under each configuration.

\vspace{.5ex}
\bsection{Metrics}
We adopted standard metrics $\mathsf{precision}$, $\mathsf{recall}$ and $F_1$ $\mathsf{score}$. A detected mode change $\widehat{\kappa}$ is a true positive ({\bf TP}) if it is within $[t_\kappa-\Delta, t_\kappa+\Delta]$ of an actual mode change at $t_\kappa$, where $\Delta=7.5$\,s. Prior work has shown that second-level accuracy like this is sufficient for practical attacks~\cite{DBLP:conf/huc/GaoFSKYL14,DBLP:conf/ipsn/KangS20}. 
To avoid multiple detections matching the same ground truth, we enforce a minimum distance of $2\Delta$ between changepoints using the $\mathsf{min\_size}$ parameter in $\mathsf{ruptures}$.
Each unmatched detection counts as a false positive ({\bf FP}), and each unmatched ground truth as a false negative ({\bf FN}). By definition, $\mathsf{precision}$ = TP/(TP+FP), $\mathsf{recall}$ = TP/(TP+FN), and $F_1 = 2\times\mathsf{precision}\times\mathsf{recall} / (\mathsf{precision}+\mathsf{recall})$. 

As defining meaningful events of interest is challenging with synthetic tasksets, we focus on evaluating the quality and overhead of mode change detection in this section.

\bsection{Attack parameters}
Mode change detection is an unsupervised process: no model is trained,
and ground-truth mode changes serve only to choose two CPD parameters. We use
the percentiles $\{0,10,25,75,90,100\}$ of each window as features and sliding
windows of $w_c=5000$ and $w_f=100$ jobs, each advancing by half its size. The
best CPD cost function and penalty $\beta$ depend on the target taskset; as in
the preparation phase of the case studies, the attacker tunes them for the
target system. We emulate this by selecting, for each taskset, the cost
function from $\{\ell_2, \text{linear}, \text{RBF}\}$ and the penalty from 1 to
20 that detect the mode change with the fewest false alarms. Since this
selection uses the same 600-s simulation that we evaluate, the results reflect
a well-tuned attacker.

\begin{figure*}[t]
    \centering
    \includegraphics[width=.85\textwidth, trim={0ex 2ex 0ex 1ex}, clip]{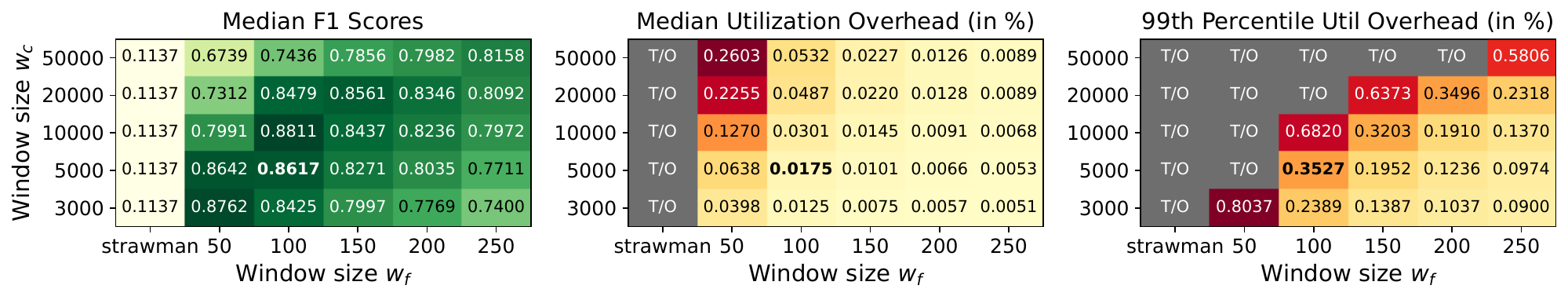}
    \vspace*{-2.75ex}
    \caption{Detection quality and CPU overhead with varying window sizes. Default configuration is {\bf bolded}.}
    \label{fig:win_analysis}
    \vspace*{-2.25ex}
\end{figure*}

\bsection{Implementation and platform}
We implemented \sys in Python 3.12.3 and ran simulations on AMD EPYC 7543 CPUs with 256\,GB RAM on $\mathsf{CloudLab}$~\cite{cloudlab}.  We assume no task migration overhead or task release jitter in the simulation.

\subsection{Detection quality} \label{subsec:eval-base}
Fig.~\ref{fig:win_analysis} shows detection quality (the $F_1$ score) and CPU utilization overhead (in \%) under varying sliding window sizes in a single-core setting with an observer period of no more than 200\,ms. A much larger observer period leads to fewer observations and thus lower detection quality. 
The figure also includes results for a baseline (`strawman') that performs changepoint detection (PELT) directly on raw data without using a fine-grained window.
For both methods, we limit detection to 5 s; a run that exceeds this limit is treated as detecting no changepoints, and overhead values beyond it are shown as T/O.
The results show that incorporating a fine-grained window significantly improves both accuracy and efficiency compared to raw PELT. 
However, as $w_f$ increases beyond a threshold, detection quality degrades, as changepoints are masked within larger windows. Overhead is highly sensitive to $w_f$, as smaller fine-grained windows result in more frequent feature extractions and longer input sequences per coarse-grained window. In contrast, the coarse-grained window $w_c$ mainly affects invocation frequency and has a weaker impact on runtime. 

\subsection{\sys with schedule-randomization defense} \label{subsec:eval-defense}
\noindent We implemented a state-of-the-art schedule-randomization defense that perturbs job inter-arrival times using Laplace noise to reduce schedule determinism~\cite{chen2021indistinguishability}. The noisy inter-arrival interval of a task $\tau_i$ is defined as $\mathsf{Lap}(T_i, f(\epsilon), T_i^{\bot}, T_i^{\top})$, where $f(\epsilon)$ denotes the noise scale as a function (see details in~\cite{chen2021indistinguishability}) of the privacy budget $\epsilon \geq 0$ (i.e., the maximum acceptable privacy loss). Smaller values of $\epsilon$ correspond to stronger privacy guarantees and correspondingly higher noise levels. 
To ensure feasibility, the perturbed inter-arrival times are constrained to remain within $[T_i^{\bot}, T_i^{\top}]$, preserving non-negative intervals and schedulability requirements. This perturbation also makes the tasks inherently \emph{sporadic}.

Following the approach
of~\cite{chen2021indistinguishability}, we considered two variants:
(i) a {\em task-level variant}, which uses a common noise scale
and common release-interval bounds $[T^\bot,T^\top]$ across all
tasks; and
(ii) a {\em restricted job-level variant}, which uses task-specific
release-interval bounds $[T_i^\bot,T_i^\top]$, while sharing a
common noise scale across tasks.

\begin{table*}[t]
\small
\centering
\scalebox{0.9}{
\begin{tabular}{|cc|c|c|c|c|c|c|}
\hline
\multicolumn{2}{|c|}{\textbf{Defense Configuration}} &
  No defense &
  $\epsilon=10$ &
  $\epsilon=10^3$ &
  $\epsilon=10^4$ &
  $\epsilon=10^5$ &
  $\epsilon=10^6$ \\ \hline
\multicolumn{1}{|c|}{\multirow{3}{*}{\textbf{Defense Effects}}} &
  Precision of mode change detection &
  0.9858 &
  0.7387 &
  0.7219 &
  0.8577 &
  0.9250 &
  0.9389 \\ \cline{2-8} 
\multicolumn{1}{|c|}{} & Recall of mode change detection                             & 0.8342 & 0.8397 & 0.7925 & 0.5451 & 0.5340 & 0.6541 \\ \cline{2-8} 
\multicolumn{1}{|c|}{} & $F_1$ score of mode change detection                        & 0.9037 & 0.7860 & 0.7556 & 0.6666 & 0.6771 & 0.7710 \\ \hline
\multicolumn{1}{|c|}{\multirow{2}{*}{\textbf{Side Effects}}} &
  Ratio of jobs completed in time, relative to no defense &
  1.0000 &
  0.1440 &
  0.1500 &
  0.2297 &
  0.5390 &
  0.8307 \\ \cline{2-8} 
\multicolumn{1}{|c|}{} & Ratio of tasksets with introduced vulnerability & N/A    & 0.1414 & 0.1130 & 0.0606 & 0.0266 & 0.0291 \\ \hline
\end{tabular}
}

\vspace{.5ex}\caption{The defense effects and side effects of task-level defense. Smaller $\epsilon$ means larger noise.}
\label{tbl:task-level-def}
\vspace*{-7ex}
\end{table*}

\vspace{.5ex}
\bsection{Results}
Table~\ref{tbl:task-level-def} reports the detection quality and job completion results for the task-level variant with
$T^{\bot}=1$~ms and $T^{\top}=2000$~ms, which cover all task periods,
and privacy budget $\epsilon$ varied in $\{10,10^3,10^4, 10^5, 10^6\}$. 
The restricted job-level variant gives similar findings
(Appendix~\ref{app-sec:synthetic}). 
Overall, noise lowers precision at every $\epsilon$ and recall at every
$\epsilon$ except $\epsilon=10$. This is expected: perturbed inter-arrival
times can create spurious changepoints, which lower precision, or smooth
out true distributional shifts, which lowers recall. 
Notably, stronger noise does not always protect better: intermediate values (e.g., $\epsilon=10^4$) can yield greater degradation than smaller values (e.g., $\epsilon=10$), indicating a non-trivial interaction between noise magnitude and the structure of schedule-induced timing signals. Further inspection suggests that randomization can, in some cases, alter interleaving patterns between the observer and other tasks in a way that increases separability of certain mode changes and increases detection quality, thereby partially offsetting the intended obfuscation effect. 
This phenomenon is illustrated in the last row of Table~\ref{tbl:task-level-def}, which shows the ratio of tasksets whose mode changes \sys could not detect when there was no defense but correctly detected under the defense.

Finally, we evaluated the defense's cost in \emph{timely job completion}, i.e.,
the number of jobs that complete by their deadlines, relative to the case
without defense. Small values of $\epsilon$ ($\epsilon \le 10^4$) reduce timely
job completion to 14\%--23\%, mainly because inter-arrival times inflate and
fewer jobs are released. Fewer timely jobs is a cost in itself: for a periodic
controller, every job that is not released or misses its deadline is a skipped
control update.

\section{Mitigation} \label{sec:discussion}
\sys exploits mode-dependent changes in when untrusted tasks run and on
which cores. Randomizing release times (Sec.~\ref{subsec:eval-defense}) does not remove this signal, because a work-conserving scheduler still exposes
differences in workload between modes to co-running tasks. Mitigation must instead make when and where untrusted tasks run independent of the
mode, analogous to constant-time code.
\emph{Spatial isolation} assigns sensitive and untrusted tasks to
separate cores with fixed affinities, and must also partition shared
caches and memory bandwidth and fix the CPU frequency.
\emph{Temporal isolation} gives sensitive and untrusted tasks separate, fixed execution windows. In both cases, capacity that sensitive tasks
leave unused must stay idle rather than go to untrusted tasks.
Hiding core identities from untrusted tasks is cheaper but weak: it
lowers \sys's  $F_1$ on four-core tasksets only from 0.93 to 0.88 
(Table~\ref{tbl:ablation}  in Appendix~\ref{app-sec:synthetic}).

\bsection{Deployment costs} 
Spatial isolation may require additional cores.
S3K~\cite{karlsson2025partitioning} supports such temporal isolation but notes
that changing these execution windows at run time can reveal modes. Fixed windows must therefore cover the sensitive tasks' maximum demand over all modes, which leaves capacity idle in lighter-load modes.  Since any task with a
vulnerable dependency can become an observer, robust protection may
require isolating individual tasks.

\section{Related Work}\label{sec:related-work}

\noindent{\bf Side-channel attacks against CPS.}
A broad range of CPS side channels has been studied~\cite{Tsalis2019}. For instance, prior work has shown that vehicle trajectories can be inferred from 
cache access patterns~\cite{luo2020stealthy}, speed signals~\cite{DBLP:conf/huc/GaoFSKYL14,DBLP:conf/icdcs/ZhouCLZC17}, brake signals~\cite{DBLP:journals/iotj/SarkerSQUZ23}, and battery levels~\cite{DBLP:conf/ipsn/KangS20}. 
In embedded and IoT systems, additional attacks exploit network traffic~\cite{IoTMosaic,DBLP:journals/corr/ApthorpeRF17}, power consumption~\cite{DBLP:journals/ieeesp/LisovichMW10}, and acoustic emanations~\cite{DBLP:conf/sp/HuLSC23,DBLP:conf/sp/LongNKBRF23} to infer private information. In contrast, \sys uses only the timing and core assignments of the observer's jobs.

\noindent{\bf Schedule-based side channels.} 
Information leakage through schedule-based side channels has been extensively studied~\cite{DBLP:conf/uss/VilaK17,kadloor2010low,DBLP:journals/ton/GongK16}. 
In real-time CPS, prior attacks use an observer task to infer a victim task's execution slot~\cite{chen2019novel,chen2020scheduler}, execution mode~\cite{babar2025investigating}, or task period~\cite{liu2019leaking}, in a \emph{single-core} system. 
These attacks assume detailed system knowledge, such as task parameters, and target a single victim task. Closest to our setting, Liu et al.~\cite{liu2019leaking} infer engine speed from the period of a rate-adaptive task, but require an explicit period-to-speed mapping. \sys instead targets multi-mode systems: it needs neither the taskset nor the scheduling policy, works on multicore platforms, and infers sensitive events, such as a vehicle's route, rather than individual task parameters.

\noindent{\bf Defenses against schedule-based attacks.}
Existing defenses fall into three classes.
The first restricts co-scheduling within an ``attack effective window'' of a
specific victim task~\cite{chen2021schedguard,chen2023schedguard++,ren2023protection,ren2024multimode,sain2024enhancingattackresiliencerealtime},
and thus does not address attacks that exploit global scheduling behavior
without a target task.
The second reduces scheduling determinism via randomization or
reordering~\cite{chen2018reorder,yoon2016taskshuffler,DBLP:conf/dsn/YoonKBS22,chen2021indistinguishability};
as our evaluation shows, the closest one~\cite{chen2021indistinguishability}
only partially mitigates \sys while degrading schedulability and introducing
new vulnerabilities (cf.~\cite{nasri2019pitfalls,babarunderstanding}).
The third enforces noninterference by filling the protected workload's unused
service with idle time, e.g., idle placeholders for blocked high-security
threads~\cite{DBLP:conf/ccs/VolpHH08} or dummy slots for private
clients~\cite{DBLP:conf/ccs/BeamsKA21}.
These can prevent \sys only if the reserved service is fixed across modes,
which requires provisioning for the maximum demand over all modes
(Sec.~\ref{sec:discussion}).

\bsection{Opacity}
Opacity studies whether an observer can determine a system's secret state or behavior from its outputs~\cite{jacob2016overview}. Approximate opacity extends this question to CPS with outputs that are indistinguishable within a measurement threshold~\cite{yin2021approximate}. 
\sys instead measures how accurately sensitive events can be inferred from noisy scheduling observations.

\section{Conclusion}
\noindent This paper shows that mode changes in real-time systems can introduce previously underexplored security vulnerabilities. We presented \sys, a lightweight attack that infers sensitive events associated with mode changes using only passive timing observations and minimal system knowledge. Across synthetic workloads and real-world case studies, \sys achieves accurate detection under different scheduling policies and system configurations. 
These results underscore the challenge of preserving real-time guarantees while mitigating timing leakage in multi-mode real-time CPS.

\clearpage

\bibliographystyle{plain}
\bibliography{paper}

\clearpage

\section*{Appendices}
\appendix

\section{Supplementary materials for the vehicle route prediction case study}\label{sec:appendix-amcl}

\begin{figure}[h]
    \centering
    \includegraphics[width=0.9\linewidth]{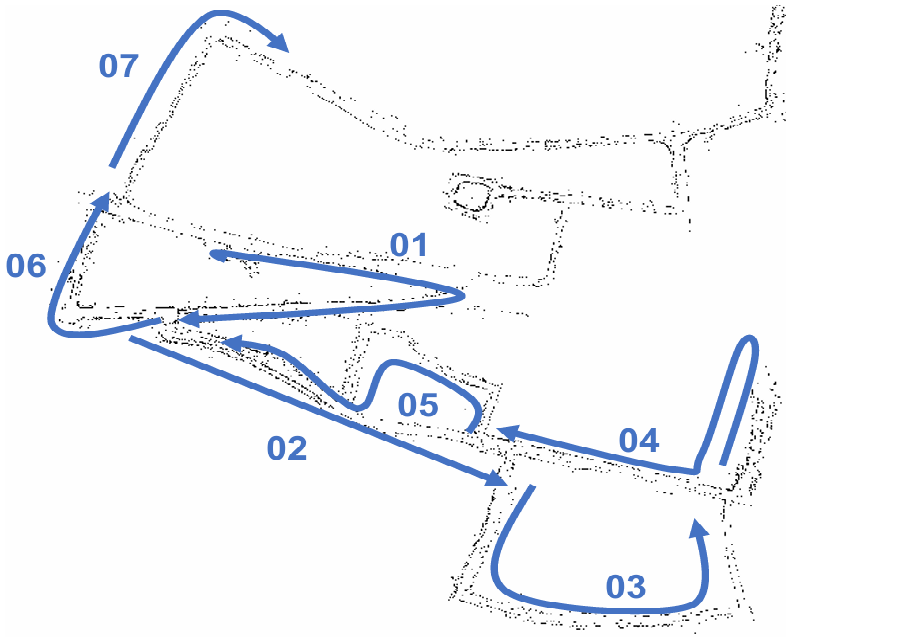}
    \caption{Map and routes for the route prediction case study.}
    \label{fig:oxford}
\end{figure}

\bsection{Route segmentation}
\noindent Fig.~\ref{fig:oxford} shows an overview of the map used in the vehicle case study (Sec.~\ref{subsec:eval-amcl}). For a fair comparison, we used the same dataset and followed the map construction and route segmentation procedure of~\cite{luo2020stealthy}, including the seven-route labeling shown in Fig.~16 of~\cite{luo2020stealthy}. 

\bsection{Details of the classifiers}
The CNN takes the window sequence $\mathbf{F}^{\mathsf{win}}$ with 1-s windows (nine features per timing signal, i.e., 18 input channels when $T_r$ is used) and applies five 1-D convolutional layers (kernel sizes 7, 5, 5, 3, 3; dilations 1, 2, 2, 4, 8; 32 to 96 filters), each followed by batch normalization and a ReLU. The first three layers halve the sequence by max-pooling, so each output of the last layer covers about 250\,s of the trace. Traces have different lengths, so each sequence is zero-padded to a common length, and a mask removes the padding after every layer. Average- and max-pooling over the unpadded part then reduce the trace to a fixed-length vector. This vector is concatenated with an embedding of the trace duration and the observer configuration, and a two-layer classifier with dropout maps it to the seven routes.
Its output is averaged over three random initializations. To combine the four classifiers (the two ExtraTrees, HGB, and the CNN), \sys weights the log-probabilities of each by a value in $\{0, 0.25, \ldots, 2\}$ and selects the four weights jointly by exhaustive search, maximizing accuracy on held-out traces within the training data. The weights are shared across all 12 configurations, except for the zero-utilization case where $T_r$ is constant, so the weights of classifiers using $T_r$ are set to zero.

\begin{figure*}
            \centering
            \includegraphics[width=0.85\textwidth,trim={0 1.5ex 0 0}, clip]{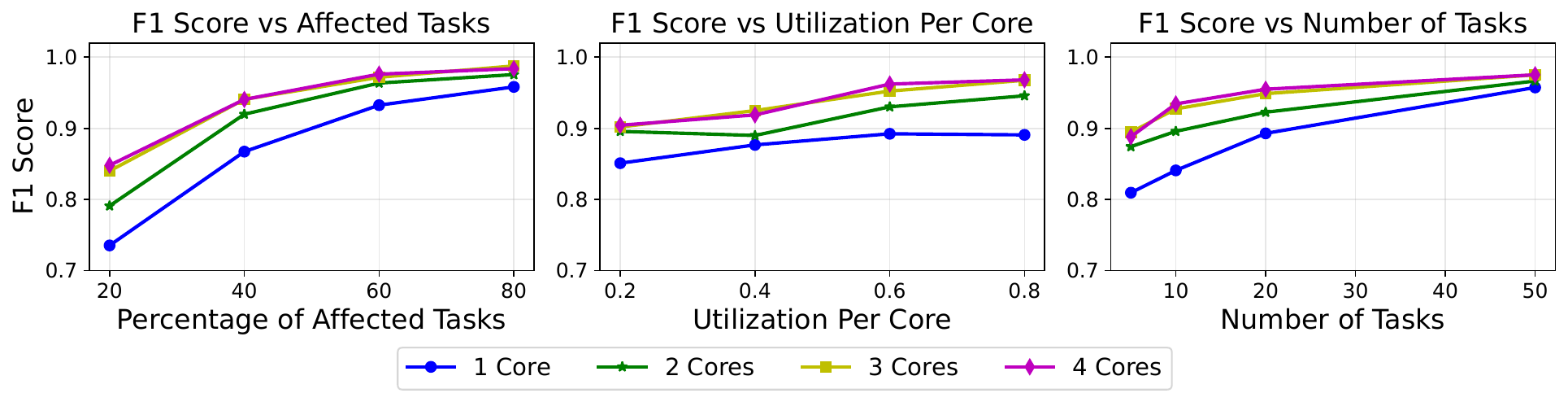}
            \vspace*{-2ex}
            \caption{Mode change detection quality ($F_1$ score) on multicore platforms.}\label{fig:f1-multicore-total-util}
            \vspace*{-2ex}
\end{figure*}

\begin{figure*}[t]
    \centering
    \includegraphics[width=.85\linewidth]{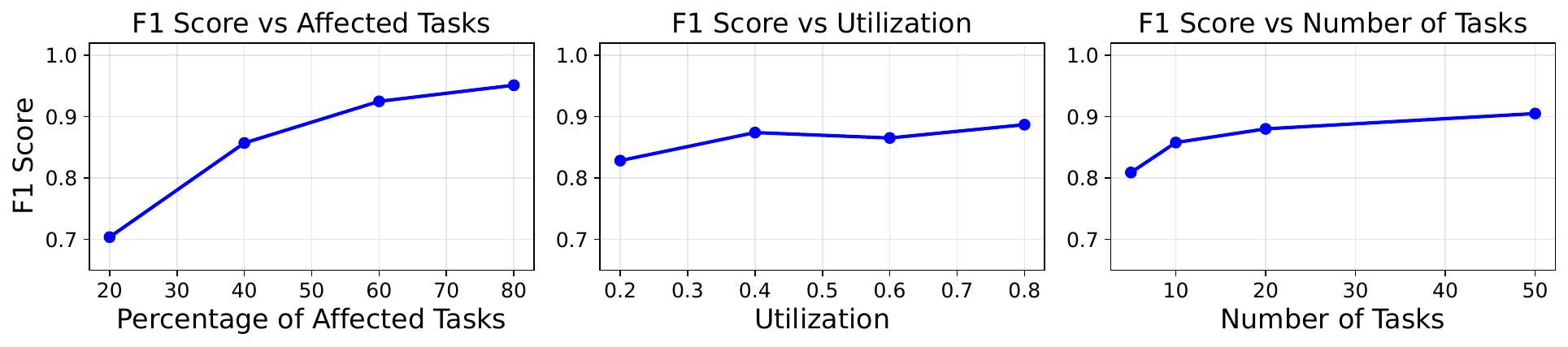}
    \caption{Detection quality of \sys under Rate Monotonic scheduling policy}
    \label{fig:rm-f1}
\end{figure*}

\begin{figure*}[t]
    \centering
    \includegraphics[width=.85\textwidth, trim={0ex 1.5ex 0ex 0ex}, clip]{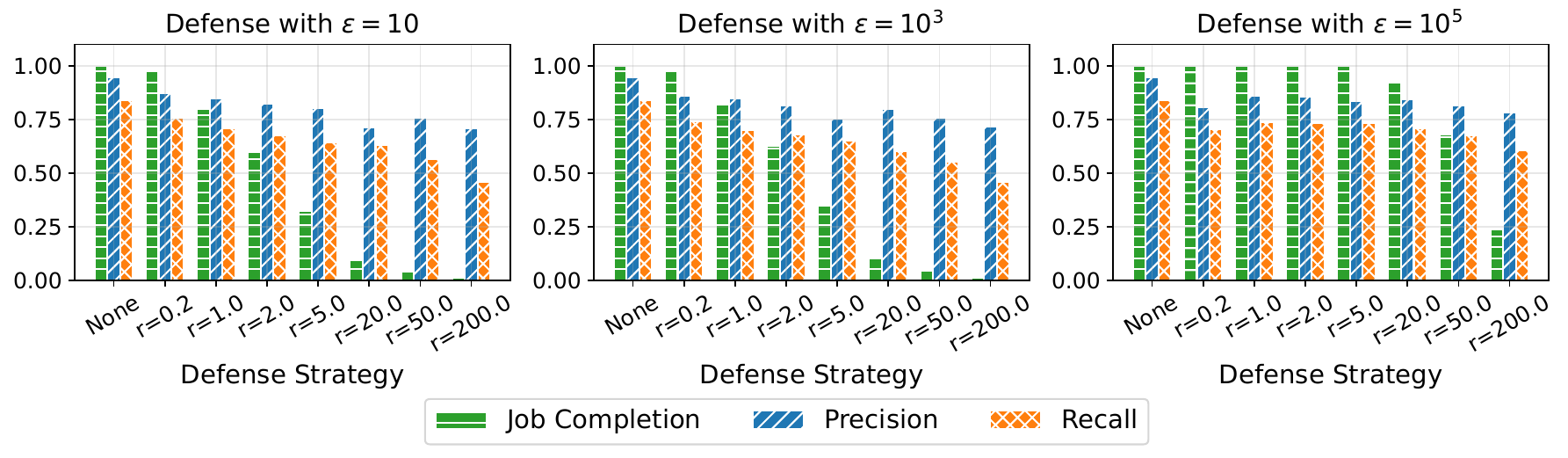}
    \caption{Detection quality of \sys and job completion rate with the restricted job-level variant.}
    \label{fig:perf-with-def}
    \vspace*{-3ex}
\end{figure*}

\section{Additional evaluation on synthetic tasksets}
\label{app-sec:synthetic}
In this section, we show results on the parameter sensitivity analysis and ablation study on the synthetic tasksets (Sec.~\ref{sec:evaluation}). 

\bsection{Results under different configurations}
Fig.~\ref{fig:f1-multicore-total-util} shows the $F_1$ scores for mode-change detection, demonstrating that \sys maintains strong detection performance across a wide range of system parameters. Detection accuracy improves as more tasks are affected by a mode change, since larger schedule perturbations are easier to distinguish (left subfigure). Performance also increases with system utilization and taskset size, as both lead to richer timing variation and more observable interference patterns (middle and right subfigures). Finally, detection is more accurate with more cores under the same normalized utilization, likely due to the higher total system load and the resulting increase in observable changes in timing patterns. 

\bsection{Results with Rate Monotonic scheduling}
We also evaluated the detection quality with the Rate Monotonic (RM) scheduling policy~\cite{DBLP:journals/jacm/LiuL73}, a standard real-time fixed-priority scheduling policy that assigns higher priorities to tasks with shorter periods. Since task periods were generated randomly, the observer’s priority under RM is also random. We used the same tasksets and evaluation metrics as in Sec.~\ref{sec:evaluation}.

Fig.~\ref{fig:rm-f1} shows the resulting detection quality of \sys under RM scheduling. The results show that \sys maintains strong performance under a different scheduling policy from EDF.

\bsection{Ablation study}
\noindent To evaluate the contribution of individual features used by \sys, we conducted an ablation study, summarized in Table~\ref{tbl:ablation}. Starting from the full element set of the traces, we removed one element at a time and measured the resulting $F_1$ score. For the single-core setting, we evaluated the impact of removing $T_s$, $T_r$, and percentile-based features. For the multi-core setting, we additionally evaluated the impact of removing the core-ID feature.

The results show that $T_s$, percentile features, and core IDs each contribute moderately to detection quality: removing them reduces the $F_1$ score by 0.03--0.07. In contrast, removing $T_r$ causes a substantially larger degradation. Further analysis shows that this effect is particularly pronounced under EDF scheduling. Since EDF priorities vary dynamically with job deadlines, the set of tasks that can preempt the observer changes across jobs. The $T_r$ feature directly captures this intra-job preemption behavior, making it especially informative of mode-dependent scheduling and execution-time variations.

\begin{table}[t]
\centering
\footnotesize
\scalebox{0.999}{
\begin{tabular}{|c|ccc|c|}
\hline
 & \multicolumn{3}{c|}{\textbf{Single-Core}} 
 & \textbf{Four-Core} \\ \hline
 & \multicolumn{1}{c|}{\textbf{w/o $T_s$}} 
 & \multicolumn{1}{c|}{\textbf{w/o $T_r$}} 
 & \textbf{w/o Percentiles} 
 & \textbf{w/o Core ID} \\ \hline
\textbf{F1 Score} 
 & \multicolumn{1}{c|}{0.84} 
 & \multicolumn{1}{c|}{0.66} 
 & 0.80 
 & 0.88 \\ \hline
\textbf{All Features} 
 & \multicolumn{3}{c|}{0.87} 
 & 0.93 \\ \hline
\end{tabular}
}
\caption{Ablation study of features used by \sys.}
\label{tbl:ablation}
\vspace*{-3ex}
\end{table}

\bsection{Against the job-level variant}
We evaluate the restricted job-level variant
(Sec.~\ref{subsec:eval-defense}) with task-specific noise bounds.
For each task $\tau_i$, we define
$T_i^\bot=T_i/(1+r)$ and $T_i^\top=T_i(1+r)$,
where $r$ controls the relative perturbation range.
We varied $\epsilon\in\{10,10^3,10^5\}$ across different values
of $r$.
For each $(\epsilon,r)$ combination, we evaluated the impact on
\sys's performance and timely job completion.
We report the number of jobs completed on time as a fraction of
the no-defense completion count over the same observation interval.
This metric captures both reduced job releases and deadline misses.

Fig.~\ref{fig:perf-with-def} shows the results. 
Smaller $\epsilon$ ($10$ and $10^3$) and larger $r$ lower \sys's detection quality, as expected from stronger perturbation. As with the task-level variant, the trend is not strictly monotonic: some configurations make   certain mode changes easier to detect due to induced changes in scheduling structure. We also observed 2.7\% to 5.5\% of the tasksets whose mode change \sys detects only under the defense. This ratio is smaller than under the task-level variant (up to 14.1\%), likely because the job-level variant bounds the noise more tightly. 

Overall, while job-level randomization can degrade \sys under sufficiently large noise, it introduces a strong trade-off with real-time feasibility. Larger values of $r$ and smaller $\epsilon$ significantly reduce job completion rates, indicating that preserving on-time job completion requires restricting the perturbation magnitude, which in turn limits the defense's effectiveness.

\end{document}